\documentclass{ws-ijmpcs}

\def\trimmarks{}

\newcommand{\open}{{<\kern -0.3 em{\scriptscriptstyle )}}}

\newcommand{\nslash}{\kern 0.2 em n\kern -0.45em /}
\newcommand{\Pslash}{\kern 0.2 em P\kern -0.56em \raisebox{0.3ex}{/}}
\newcommand{\pslash}{\kern 0.2 em p\kern -0.4em /}
\newcommand{\kslash}{\kern 0.2 em k\kern -0.45em /}
\newcommand{\Sslash}{\kern 0.2 em S\kern -0.56em \raisebox{0.3ex}{/}}

\newcommand{\eq}{\begin{equation}}
\newcommand{\ee}{\end{equation}}
\newcommand{\beq}{\begin{equation}}
\newcommand{\eeq}{\end{equation}}
\newcommand{\ba}{\begin{eqnarray}}
\newcommand{\ea}{\end{eqnarray}}

\newcommand{\sumint}{\kern 0.2 em {\textstyle\sum} \kern -1.1 em \int}

\newcommand{\simorderr}{\raisebox{-4pt}{$\, \stackrel{\textstyle <}{\sim} \,$}}

\begin{document}

\markboth{Dani\"el Boer}
{Overview of TMD evolution}

%
\catchline{}{}{}{}{}
%

\title{Overview of TMD evolution}

\author{Dani\"el Boer}

\address{Van Swinderen Institute, University of Groningen\\
Nijenborgh 4, NL-9747 AG Groningen, The Netherlands\\
d.boer@rug.nl}

\maketitle

\begin{history}
\received{Day Month Year}
\revised{Day Month Year}
\published{Day Month Year}
\end{history}

\begin{abstract}
Transverse momentum dependent parton distributions (TMDs) appear in many scattering processes at high energy, from the semi-inclusive DIS experiments at a few GeV to the Higgs transverse momentum distribution at the LHC. Predictions for TMD observables crucially depend on TMD factorization, which in turn determines the TMD evolution of the observables with energy. In this contribution to SPIN2014 TMD factorization is outlined, including a discussion of the treatment of the nonperturbative region, followed by a summary of results on TMD evolution, mostly applied to azimuthal asymmetries.    
\end{abstract}

\ccode{PACS numbers: 12.38.-t; 13.88.+e}

\section{TMD factorization}	
Many angular asymmetries in the transverse momentum distribution of produced hadrons in semi-inclusive deep inelastic scattering (SIDIS),   
$e\, p \to e'\, h\, X$, have been measured by the HERMES, COMPASS, and JLab experiments. Evolution is needed to compare those results obtained at different energies. The evolution is dictated by the appropriate factorization. As SIDIS is sensitive to the transverse momentum of quarks through a measurement of $P_{h\perp}$, the observed transverse momentum of the produced hadron, it is naturally described within the framework of TMD factorization. 
Several forms of TMD factorization have been put forward in the literature for a number of processes 
[\refcite{CS81}-\refcite{MWZ}], which besides SIDIS includes the Drell-Yan (DY) process (lepton pair production in hadron-hadron collisions), back-to-back hadron production in electron-positron annihilation ($e^+e^- \to h_1 h_2 X$), and Higgs production. The main differences among the various approaches concern the treatment of spurious rapidity or lightcone divergences, in order to make each factor well-defined, and the redistribution of contributions to avoid the appearance of large logarithmic corrections. 
For brief summaries and comparisons cf.\ 
[\refcite{Collins:2011ca}-\refcite{Collins:2014jpa}]. 
Schematically the TMD factorization is of the form [\refcite{Collins:2011zzd}]:
\beq
d \sigma = {H \times \text{convolution of}\ A\, B} + \text{high-}q_T\ \text{correction}\ (Y)+ \text{power-suppressed}
\eeq
Here $A$ and $B$ are TMD parton distribution or fragmentation functions and $H$ is the partonic hard scattering factor. 
A soft factor has been absorbed into $A$ and $B$ [\refcite{Collins:2011zzd,EIS2012}]. The convolution in terms of $A$ and $B$ is best 
deconvoluted by Fourier transform. More specifically, for SIDIS the differential cross section is given by:
\beq 
\frac{d\sigma}{dx dy dz d\phi {d^2 \bm{q}_{T}^{}}} = \int d^2 {b} \, e^{-i {\bm{b} \cdot \bm{q}_T^{}}} \tilde{W}({\bm{b}}, Q;
  x, y, z) + {\cal O}\left(Q_T^2/Q^2\right).
\eeq
The correction is relevant at large $Q_T\, (= P_{h\perp}/z)$ and is commonly referred to as the $Y$-term. 
For unpolarized hadrons and quarks of flavor $a$, $\tilde W$ consists of 3 factors:
\beq
\tilde{W}({\bm{b}}, Q; x,y,z) =
{\sum_{a}\,\tilde{f}_1^{a}(x,{\bm{b}};\zeta_F, \mu)}
{\tilde{D}_1^{a}(z,{\bm{b}};\zeta_D,\mu)} H\left(y,Q;\mu\right).
\eeq
The Fourier transforms $\tilde{f}_1$ and $\tilde{D}_1$ of the unpolarized TMD distribution and fragmentation functions, are functions of the momentum fraction $x$ or $z$,  
transverse coordinate $\bm{b}$, rapidity variable $\zeta$, and renormalization scale $\mu$. Here 
$\zeta_F = M^2 x^2 e^{2(y_{P}-y_s)}$ and $\zeta_D = M_{h}^2 e^{2(y_s-y_{h})}/z^2$, where $y_s$ is an arbitrary rapidity that drops out of the final answer and 
$\zeta_F \zeta_D \approx Q^4$, with $Q$ the hard scale. 
The operator definition of the TMDs involves a gauge link or Wilson line $U$, which arises from   
summation of all insertions of gluons with longitudinal polarization that are not power suppressed. The path of the Wilson lines depends on whether 
the color flow in the process is incoming or outgoing 
[\refcite{Collins:1983pk}-\refcite{BMP}].
This does not automatically imply that observables depend on this path, but it does in certain cases, 
for example, the Sivers asymmetries [\refcite{Brodsky:2002rv,Collins:2002kn}], where the transverse momentum dependence is correlated with the proton spin direction [\refcite{Sivers:1989cc}]. 
The more hadrons are observed in a process, the more complicated the color flow, leading to more complicated expressions 
[\refcite{Bomhof:2006dp,Buffing:2013dxa}] or sometimes even factorization breaking 
[\refcite{Collins:2007nk}-\refcite{Rogers:2010dm}]. In addition, the gauge links in SIDIS and DY have lightlike pieces which lead to spurious lightcone divergences. As a regularization, 
the path can be taken off the lightcone, specified by some finite rapidity. The variation in this rapidity determines the change of the TMD 
with $\zeta$. Also, the regularization allows for calculation of the Sivers and Boer-Mulders effects on the lattice [\refcite{Musch:2011er}]. 

Choosing the renormalization scale $\mu = Q$ avoids large logarithms in the hard scattering part $H$, but generates them in the TMDs. 
For this reason one usually evolves the TMDs to the scale 
$\mu_b=C_1/b=2e^{-\gamma_E}/b$ $(C_1 \approx 1.123)$ [\refcite{Collins:2011zzd}].
This can be done using the Collins-Soper and renormalization group equations:
\begin{align}
& \frac{d \ln \tilde{f}(x,b;\zeta,\mu)}{d \ln \sqrt \zeta} = \tilde K(b;\mu),\qquad \frac{d \ln \tilde{f}(x,b;\zeta,\mu)}{d\ln \mu}  = \gamma_F(g(\mu);\zeta/\mu^2),
\end{align}
with $d\tilde K/d\ln \mu  = -\gamma_K(g(\mu))$ \& $ \gamma_F(g(\mu);\zeta/\mu^2)  = \gamma_F(g(\mu);1)-\frac{1}{2}\gamma_K(g(\mu))\ln(\zeta/\mu^2)$. 
Using these equations one can evolve the TMDs to the scale $\mu_b$:
\beq
\tilde{f}_1^a(x,b^2; \zeta_F, \mu) \, 
\tilde{D}_1^b(z,b^2;\zeta_D, \mu) = e^{-S(b,Q)} \tilde{f}_1^a(x,b^2; \mu_b^2,\mu_b) \, 
\tilde{D}_1^b(z,b^2; \mu_b^2,\mu_b),
\eeq
where the Sudakov factor for $\zeta_F=\zeta_D=\mu=Q$ is given by
\beq
S(b,Q) = - \ln\left(\frac{Q^2}{\mu_b^2}\right)\tilde K(b,\mu_b) - \int_{\mu_b^2}^{Q^2} \frac{d\mu^2}{\mu^2} \big[ \gamma_F(g(\mu);1)-\frac{1}{2} \ln\left(\frac{Q^2}{\mu^2}\right) \gamma_K(g(\mu))\big].
\eeq
The perturbative expression for the Sudakov factor can be used whenever the restriction $b^2 \ll 1/\Lambda^2$ is justified (e.g.\ at very large $Q^2$).
If also contributions at larger $b$ are important, e.g.\ at moderate $Q$ and small $Q_T$, 
then one needs to include a nonperturbative Sudakov factor $S_{NP}$, for instance as follows: $\tilde{W}(b) \equiv \tilde{W}(b_*) \, e^{-{S_{NP}(b)}}$, 
where $b_*=b/\sqrt{1+b^2/b_{\max}^2} \leq b_{\max}$. For $b_{\max} = 1.5\ {\rm GeV}^{-1}$, $\alpha_s(C_1/b_{\max})\approx 0.6$, such that $W(b_*)$ 
can be calculated within perturbation theory. In general the nonperturbative Sudakov factor is $Q$ dependent and of the form [\refcite{Collins:1985kw,CSS-85}]:
${S_{NP}(b,Q)} = {\ln(Q^2/Q_0^2)}{g_1(b)} + {g_A(x_A,b)} + {g_B(x_B,b)}$, where $Q_0=1/b_{\max}$ and $g_{1/A/B}$ need to be fitted to data.
Until recently $S_{NP}$ was typically chosen to be Gaussian, but it appears hard to find one universal Gaussian $S_{NP}$ that describes both 
SIDIS and DY/$Z$ production data [\refcite{Sun:2013dya}]. Different $b$ dependences are considered in [\refcite{Collins:2013zsa,Collins:2014jpa,Su:2014wpa}].

The TMDs at initial $\mu_i, \zeta_i$ and final $\mu_f, \zeta_f$ can be related by an evolutor $\tilde R$, i.e.\
$\tilde{f}(x,b; \zeta_f, \mu_f) = \tilde R(b;\zeta_i,\mu_i,\zeta_f,\mu_f) \tilde{f}(x,b; \zeta_i, \mu_i)$, with
\beq
\tilde R(b;\zeta_i,\mu_i,\zeta_f,\mu_f) = \exp\left\{\int_{\mu_i}^{\mu_f} \frac{d\bar\mu}{\bar\mu}
\gamma_F\left(\alpha_s(\bar\mu),\ln\frac{\zeta_f}{\bar\mu^2} \right)
\right\} \left( \frac{\zeta_f}{\zeta_i} \right)^{-D\left(b;\mu_i\right)}.
\eeq
In [\refcite{Echevarria:2012pw}] resummation of logarithms in the perturbative expression for $D(b,\mu)=-\frac{1}{2}\tilde{K}(b,\mu)$ is performed to NNLL 
order. To this order the resummed expression $D^R$ shows convergence for $b \simorderr b_X/2$, where to leading order 
$b_X = \frac{C_1}{\mu_i} \exp \left(\frac{2\pi}{\beta_0 \alpha_s(\mu_i)}\right)$ [\refcite{Echevarria:2012pw}]. 
The resummed evolutor $\tilde R$ vanishes well before $b \sim b_X/2$ if $\mu_f \gg \mu_i$ and may thus reduce the impact of the nonperturbative 
$b$ region. Using the $b_*$ method this approach favors $b_{\max} \sim 1.5\ {\rm GeV}^{-1}$ [\refcite{Echevarria:2012pw}]. 
Similar resummations in the perturbative expansion of the TMDs are performed in [\refcite{D'Alesio:2014vja}]. 
It shows that at low scales the TMDs are very small at large $b$ where $\alpha_s(\mu_b)$ is very large.  
Furthermore, in [\refcite{D'Alesio:2014vja}] it is suggested that the sensitivity to the Landau pole is minimized by using 
as initial scale $Q_0+q_T$ rather than $\mu_b$. Correspondingly a nonperturbative factor with a new form is considered:
$e^{-\lambda_1 b}\left(1+\lambda_2 b^2\right)\left(Q^2/Q_0^2\right)^{-\frac{\lambda_3}{2} b^2}$. High $Q$ data (DY/$Z$) need only 
$\lambda_1$ and $\lambda_2$. Low $Q$ data (SIDIS) require modification by including $\lambda_3$. 
 
\section{TMD evolution}
Under TMD evolution the shape of the transverse momentum distributions changes. 
Typically TMDs become broader and decrease in magnitude with increasing energy 
[\refcite{B01}-\refcite{APR}].
If one starts out with an approximately Gaussian distribution at low scales, then a power law tail develops under TMD evolution, see e.g.\ [\refcite{AR}].
This is in contrast to a DGLAP evolution of $f(x,k_T;\mu) \propto f(x;\mu)$ that has sometimes been considered.  
In [\refcite{Anselmino:2012aa}] it was shown that in the limited range of $Q$ from 1.5 to 4.5 GeV, where $S_{NP}$ dominates the evolution, such DGLAP evolution hardly modifies the TMDs, whereas TMD evolution reduces them by about a factor of 2.

TMD evolution has been studied for various azimuthal asymmetries, which generally decrease as the energy increases. 
The Sivers asymmetry in SIDIS and DY has been studied in 
[\refcite{Idilbi:2004vb}-\refcite{Echevarria:2014xaa,Sun:2013dya}];
the Collins effect in $e^+ e^-$ annihilation and SIDIS in [\refcite{B01,B09,Echevarria:2014rua}]; and the Sivers effect in $J/\psi$ production in [\refcite{Godbole:2013bca,Godbole:2014tha}].                                                      
The main differences among the approaches are in the treatment of the nonperturbative Sudakov factor and the treatment of leading logarithms, i.e.\ the level of perturbative accuracy. 

First we discuss the TMD evolution of the Sivers asymmetry. The HERMES data ($\langle Q^2 \rangle \sim 2.4\ {\rm GeV}^2$) lie mostly above the 
COMPASS data ($\langle Q^2 \rangle \sim 3.8\ {\rm GeV}^2$) [\refcite{Adolph:2014zba}]. As can be seen in the study of [\refcite{APR}], TMD evolution from the HERMES to COMPASS energy scale seems to work well. This result is obtained with some approximations that should be applicable at small $Q$: 1) the $Y$ term is dropped (or equivalently the perturbative tail of the TMDs);  2) evolution from a fixed starting $Q_0$ rather than $\mu_b$; 
3) Gaussian TMDs at the starting scale $Q_0$ are adopted. 
It has been observed in [\refcite{Boer:2013zca}] that under these approximations plus the assumption that the TMDs as functions of $b_*$ are slowly varying functions of $b$ in the dominant $b$ region, the $Q$ dependence of the Sivers asymmetry just resides in an overall factor: $A_{UT}^{\sin(\phi_h-\phi_S)} \propto {\cal A}(Q_T,Q)$.
Using these approximations the peak of the Sivers asymmetry decreases as $1/Q^{0.7\pm 0.1}$ and the peak of the asymmetry shifts slowly towards higher $Q_T$ [\refcite{Boer:2013zca}]. 
In [\refcite{APR}] it was found that the asymmetry {\it integrated} over the measured $x, z, P_{h\perp}$ range falls off faster than $1/Q$ but slower than $1/Q^2$. 
Testing these features needs a large $Q$ range, requiring a high-energy Electron-Ion Collider (EIC).
At low $Q^2$ (up to $\sim 20\ {\rm GeV}^2$), the $Q^2$ evolution is dominated by $S_{NP}$ [\refcite{Anselmino:2012aa}]. Precise low $Q^2$ data can help to determine the form and size of $S_{NP}$, which is responsible for the $\pm 0.1$ in $1/Q^{0.7\pm 0.1}$. CLAS12 is projected to have very precise data between 1 and 7 ${\rm GeV}^2$ (see page 32 of [\refcite{Contalbrigo}]). 

Next we discuss the TMD evolution of Collins asymmetries. The Collins effect is described by a TMD fragmentation function [\refcite{Collins:1992kk}], giving rise to a $\sin(\phi_h+\phi_S)$ asymmetry in SIDIS, in combination with the transversity TMD. Unlike the Sivers asymmetry, for the Collins asymmetry no clear need for TMD evolution from HERMES to COMPASS (2010) data is apparent. This also needs to be investigated further using future data from JLab 12 and possibly EIC. 
The Collins fragmentation function can be measured independently through the double Collins effect (DCE) $\cos 2\phi$ asymmetry in $e^+e^- \to h_1 \,h_2\, X$ [\refcite{Boer:1997mf}], which has been clearly observed by BELLE [\refcite{Abe:2005zx,Seidl:2008xc}], BaBar [\refcite{TheBABAR:2013yha}] and BESIII [\refcite{SPIN2014}].
Under similar assumptions as for the Sivers asymmetry, also the DCE asymmetry (and its double ratio for unlike sign over like sign hadron pairs) is proportional to an overall factor ${\cal A}^{\rm DCE}(Q_T)$ (cf. [\refcite{Boer:2013pya}]). It shows a considerable Sudakov suppression $\sim 1/Q$ [\refcite{B01,B09,Boer:2013pya}], which is in rough agreement with the results in 
[\refcite{Sun:2013dya,SPIN2014}]. The $1/Q$ behavior should modify the Collins effect based transversity extraction, when full TMD evolution is implemented. Due to the lower $Q$ of the BES data, here one does have to worry about $1/Q^2$ corrections (analogue of the Cahn effect) [\refcite{Berger:1979xz,Brandenburg:1994wf}], which can be bounded by studying simultaneously the $1/Q \cos\phi$ asymmetry as explained in [\refcite{B09}]. 

Finally, we turn to the Higgs transverse momentum distribution, which is also a TMD factorizing process. 
The hard scale $Q=M_H$ is fixed in this case, but TMD evolution matters nevertheless, as the gluon
TMDs are probed over a whole range of scales $\mu_b$. The Higgs transverse momentum 
distribution is sensitive to the linear polarization of gluons inside the unpolarized protons 
[\refcite{Mulders:2000sh}-\refcite{Boer:2011kf}]. It requires nonzero gluon transverse momentum, 
but unlike the Sivers and Collins TMDs, it is even in $k_T$.
Numerical studies of the effects of linear gluon polarization on the Higgs transverse momentum distribution vary 
from permille level [\refcite{Wang:2012xs}] to several percent [\refcite{Boer:2014tka}], but with quite some uncertainty 
from the nonperturbative large-$b$ region and to a lesser extent from the perturbative very small $b$ region ($b \ll 1/Q$). 
At the Higgs mass scale the uncertainty from the latter region is estimated to be less than 15\% by 
adopting different regularizations (as in [\refcite{Boer:2014tka}]), like the standard one of [\refcite{Parisi:1979se}]. 
The treatment of the very small $b$ behavior of $\tilde{W}$ becomes increasingly relevant for smaller $Q$ values (scalar quarkonium production), 
and is connected to the treatment of the $Y$-term, in order to reproduce the correct integrated cross section, which itself should not be affected 
by the linear gluon polarization. This requires further study.

\section{Summary}
There have been many significant developments on TMD factorization and evolution recently: new TMD factorization expressions without explicit soft factor and with each factor well-defined have been obtained for several important processes; additional resummations have been performed; and, there has been progress towards describing SIDIS, DY, and $Z$ production data by a universal non-perturbative function. TMD evolution has been studied (at varying levels of accuracy) for Sivers and (single and double) Collins effect asymmetries and for Higgs production including the effects of linear gluon polarization. Future data from JLab 12 and BES and perhaps a high-energy EIC can help to 
map out the $Q$ dependence of Sivers and Collins asymmetries in greater detail. Future data from LHC on Higgs (and heavy quarkonium) production and a high-energy EIC could do the same for gluon dominated TMD processes. TMD (non-)factorization at next-to-leading twist remains entirely unexplored, but the $Q^2$ 
dependence of azimuthal asymmetries at twist-3 (e.g.\ $A_{LU}^{\sin \phi}$) will be measured in detail at CLAS12, posing a remaining theory challenge. 
 

\end{document}